\pdfoutput=1

\documentclass[11pt,twoside,a4paper,cmspaper,final,collab]{cms-tdr}
\def\svnVersion{388365}\def\cmsCernNoTag{CERN-EP/2016-216}\def\cmsCernDate{\today}\def\cmsMessage{Published in Physics Letters B as \href{http://dx.doi.org/10.1016/j.physletb.2017.01.027}{\doi{10.1016/j.physletb.2017.01.027.}}}

\begin{document}\cmsNoteHeader{EXO-16-027}

\hyphenation{had-ron-i-za-tion}
\hyphenation{cal-or-i-me-ter}
\hyphenation{de-vices}
\RCS$Revision: 375327 $
\RCS$HeadURL: svn+ssh://svn.cern.ch/reps/tdr2/papers/EXO-16-027/trunk/EXO-16-027.tex $
\RCS$Id: EXO-16-027.tex 375327 2016-12-01 03:20:19Z alverson $
\newlength\cmsFigWidth
\newlength\cmsSmallFigWidth
\newlength\cmsFigMedWidth
\newlength\cmsFigDoubleWidth
\ifthenelse{\boolean{cms@external}}{\setlength\cmsSmallFigWidth{0.85\columnwidth}}{\setlength\cmsSmallFigWidth{0.38\textwidth}}
\ifthenelse{\boolean{cms@external}}{\setlength\cmsFigWidth{0.98\columnwidth}}{\setlength\cmsFigWidth{0.49\textwidth}}
\ifthenelse{\boolean{cms@external}}{\setlength\cmsFigMedWidth{0.98\columnwidth}}{\setlength\cmsFigMedWidth{0.75\textwidth}}
\ifthenelse{\boolean{cms@external}}{\setlength\cmsFigDoubleWidth{0.98\columnwidth}}{\setlength\cmsFigDoubleWidth{0.9\textwidth}}
\ifthenelse{\boolean{cms@external}}{\providecommand{\cmsLeft}{top\xspace}}{\providecommand{\cmsLeft}{left\xspace}}
\ifthenelse{\boolean{cms@external}}{\providecommand{\cmsRight}{bottom\xspace}}{\providecommand{\cmsRight}{right\xspace}}
\ifthenelse{\boolean{cms@external}}{\providecommand{\cmsULeft}{top\xspace}}{\providecommand{\cmsULeft}{upper left\xspace}}
\ifthenelse{\boolean{cms@external}}{\providecommand{\cmsURight}{middle\xspace}}{\providecommand{\cmsURight}{upper right\xspace}}

\providecommand{\PX}{\ensuremath{\mathrm{X}}\xspace}
\providecommand{\ktild}{\ensuremath{\tilde{k}}\xspace}
\providecommand{\sqrts}{\ensuremath{\sqrt{s}}\xspace}
\providecommand{\mgg}{\ensuremath{m_{\gamma\gamma}}\xspace}
\providecommand{\gaga}{\ensuremath{\gamma \gamma}\xspace}
\providecommand{\mX}{\ensuremath{m_{\PX}}\xspace}
\providecommand{\GammaX}{\ensuremath{\Gamma_{\PX}}\xspace}
\providecommand{\GammaOm}{\ensuremath{\Gamma_{\PX} / m_{\PX}}\xspace}
\providecommand{\scEta}{\ensuremath{\abs{\eta_{\text{C}}}}\xspace}
\providecommand{\RS}{{\textsc{RS}}\xspace}
\providecommand{\MGAMC}{{{\textsc{MadGraph5}}\_aMC@NLO}\xspace}
\providecommand{\mylumi}{\ensuremath{12.9\fbinv}\xspace}
\providecommand{\CLs}{\ensuremath{\mathrm{CL_{s}}\xspace}
\providecommand{\CL}{CL\xspace}}
\providecommand{\bon}{3.8\unit{T}}
\providecommand{\boff}{0\unit{T}}
\providecommand{\fb}{\unit{fb}}

\cmsNoteHeader{EXO-16-027}
\title{Search for high-mass diphoton resonances in proton-proton collisions
at 13\TeV and combination with 8\TeV search}

\date{\today}

\abstract{
  A search for the resonant production of high-mass photon pairs is presented.
  The search focuses on
  spin-0 and spin-2 resonances with masses
  between 0.5 and 4.5\TeV,
  and with widths, relative to the mass, between $1.4\times10^{-4}$ and $5.6\times10^{-2}$.
  The data sample corresponds to an integrated luminosity of 12.9\fbinv
  of proton-proton collisions collected with the CMS detector in
  2016 at a center-of-mass energy of 13\TeV.
  No significant excess is observed relative to the standard model expectation.
  The results of the search are combined statistically with those previously obtained
  in 2012 and 2015 at $\sqrt{s}=8$ and 13\TeV,
  respectively,
  corresponding to integrated luminosities of 19.7 and 3.3\fbinv,
  to derive exclusion limits on scalar resonances produced through gluon-gluon fusion,
  and on Randall--Sundrum  gravitons.
  The lower mass limits for Randall--Sundrum gravitons range from 1.95 to 4.45\TeV for
  coupling parameters between 0.01 and 0.2.  These are the most stringent
  limits on Randall--Sundrum graviton production to date.
}

\hypersetup{%
pdfauthor={CMS Collaboration},%
pdftitle={Search for high-mass diphoton resonances in proton-proton collisions at 13 TeV and combination with 8 TeV search},%
pdfsubject={CMS},%
pdfkeywords={CMS, LHC, extra dimensions, Randall-Sundrum, heavy resonance, spin-0,
  proton-proton collision, diphoton}
}

\maketitle

\section{Introduction}

\label{sec:intro}

The standard model (SM) of particle physics has been highly
successful in describing physical phenomena
but it is widely considered to be an incomplete theory
because of various shortcomings.
In particular, the SM suffers from the so-called hierarchy
problem~\cite{Sakai:1981gr}, which refers to the
large difference between the Higgs boson mass of
125\GeV~\cite{lhc-higgs-mass-comb} and the highest energy scale up to which the SM must be
valid.
Many extensions to the SM have been proposed to address the hierarchy problem, including theories
with additional space-like
dimensions~\cite{Randall:1999ee} and models with
extended Higgs boson sectors~\cite{Branco:2011iw}.
Some of these extensions predict new resonances that decay to a diphoton final state.
For example,
the Randall--Sundrum (\RS) approach~\cite{Randall:1999ee,Randall:1999vf} to extra dimensions
postulates massive excitations of spin-2 gravitons
that can decay to two photons.
A simple extension of the SM Higgs boson sector consists of the addition of a
doublet of complex scalar fields.
In such models~\cite{Craig:2013hca},
some of these additional scalar resonances
can decay to a photon pair~\cite{Dev:2014yca}.
According to the Landau--Yang theorem, the spin of a resonance decaying to two photons
can only be zero or an integer larger than one~\cite{Landau,PhysRev.77.242}.

Recently, the ATLAS and CMS Collaborations at the CERN LHC presented
results on searches for high-mass diphoton resonances in proton-proton
({\Pp\Pp}) collisions at a center-of-mass energy
of 13\TeV~\cite{altas-dipho-2015,cms-dipho-2015}.
The results were based on data collected in 2015,
corresponding to integrated luminosities of approximately 3\fbinv per experiment.
The CMS results included a combined analysis with {\Pp\Pp} collision
data at $\sqrt{s}=8\TeV$
collected in 2012~\cite{CMS-dipho-8TeV}
corresponding to an integrated luminosity of 19.7\fbinv.
Both collaborations reported the observation of a moderate excess of events
compared to SM expectations, compatible with the production of a
new resonance with a mass around 750\GeV.

In this Letter,
we report on an updated search for spin-0 resonances and \RS gravitons produced in pp
collisions and decaying to two photons.
The data were collected in 2016 with the CMS detector at
$\sqrt{s}=13\TeV$ and correspond to an integrated luminosity of
12.9\fbinv.
The analysis procedures are very similar to those presented
in Ref.~\cite{cms-dipho-2015} for the 2015 data.
A combined analysis of the 8\TeV data set of Ref.~\cite{CMS-dipho-8TeV}, the 13\TeV data
set of Ref.~\cite{cms-dipho-2015}, and the 13\TeV data set examined here is performed
to improve the sensitivity of the results.
Earlier LHC searches for \RS gravitons are presented in
Refs.~\cite{ATLAS-dipho-RS-8TeV,ATLAS-dipho-RS-7TeV,ATLAS-dilep-8TeV,ATLAS-dilep-7TeV,
ATLAS-dibos-7TeV,ATLAS-dibos-WZ-7TeV,CMS-dipho-8TeV,CMS-dipho-RS-7TeV,CMS-dilep-8TeV,
CMS-dilep-7-8TeV,CMS-dijet-13TeV,CMS-dijet-8TeV,Khachatryan:2016ecr,CMS-qbh-8TeV,
CMS-dibos-boost-8TeV,CMS-dibos-7TeV,CMS-dibos-ZZ-7TeV},
and for \mbox{spin-0} particles decaying to two photons in
Refs.~\cite{ATLAS-dipho-scalar-8TeV,CMS-dipho-8TeV}.
These earlier searches are based on {\Pp\Pp} collisions
at either $\sqrt{s}=7$ or 8\TeV.

\section{The CMS detector}

The central feature of the CMS apparatus is a superconducting solenoid
of 6\unit{m} internal diameter,
providing a magnetic field of 3.8\unit{T}.
Within the solenoid volume are a silicon pixel and strip tracker,
a lead tungstate crystal electromagnetic calorimeter (ECAL),
and a brass and scintillator hadron calorimeter (HCAL).
The tracking detectors cover the pseudorapidity range $\abs{\eta}<2.5$.
The ECAL and HCAL,
each composed of a barrel and two endcap sections,
cover $\abs{\eta}<3.0$,
with the boundary between the barrel and endcaps at around $\abs{\eta}=1.5$.
Forward calorimeters extend the coverage to $\abs{\eta}<5.0$.
The ECAL consists of 75\,848 lead tungstate crystals. The barrel section
has a granularity $\Delta\eta\times\Delta\phi=0.0174{\times}0.0174$,  with $\phi$ the
azimuthal angle, while the endcap sections have a granularity that coarsens
progressively up to $\Delta\eta\times\Delta\phi=0.05{\times}0.05$.
Preshower detectors consisting of two planes of silicon sensors interleaved with
a total of $3 X_0$ of lead are located in front of the endcap sections.
Muons are measured within $\abs{\eta}<2.4$ by gas-ionization detectors
embedded in the steel flux-return yoke outside the solenoid.
A more detailed description of the CMS detector,
together with a definition of the coordinate system
and the relevant kinematic variables,
can be found in Ref.~\cite{Chatrchyan:2008zzk}.

In the barrel section of the ECAL,
for photons with energies on the scale of tens of \GeV,
an energy resolution of about 1\% is achieved for
unconverted photons and for photons that convert ``late'',
i.e., just before entering the ECAL.
The remaining barrel photons have an energy resolution of
about 1.3\% up to $\abs{\eta}=1.0$,
rising to about 2.5\% for $\abs{\eta}=1.4$.
In the endcaps,
the corresponding resolution for unconverted and late-converting photons is about 2.5\%,
while the remaining endcap photons have a
resolution between 3\% and 4\%~\cite{CMS:EGM-14-001}.

The particle-flow algorithm~\cite{CMS-PAS-PFT-09-001,CMS-PAS-PFT-10-001}
reconstructs and identifies each individual particle with an optimised combination of
information from the various elements of the CMS detector. Particle candidates are
classified as either muons, electrons, photons, $\tau$ leptons, charged hadrons, or
neutral hadrons.

A two-stage trigger system selects events of interest for the analysis.
The level-1 trigger,
composed of custom hardware processors,
selects events at a maximum readout rate of about 100\unit{kHz}
using information from the calorimeters and muon detectors.
The high-level trigger software algorithms use the
full event information to reduce the event rate to less than 1\unit{kHz}
before data storage.

\section{Event simulation}

The \PYTHIA\,8.2~\cite{Sjostrand:2014zea} event generator
with NNPDF2.3~\cite{Ball:2012cx} parton distribution functions (PDFs)
is used to produce simulated signal samples of spin-0 and spin-2 resonances decaying to two photons.
The samples are generated at leading order (LO),
with values of the resonance mass \mX in the range $0.5<\mX<4.5\TeV$.
Three values of the relative width \GammaOm are used as benchmarks:
$1.4\times10^{-4}, 1.4\times10^{-2}$, and $5.6\times10^{-2}$,
where \GammaX is the width of the resonance.
These relative widths correspond, respectively,
to resonances much narrower than, comparable to,
and significantly wider than the detector resolution.
In the context of the \RS graviton model,
for which $\GammaOm=1.4\,\ktild^2$~\cite{Davoudiasl:1999jd},
the relative widths correspond to the dimensionless coupling parameter $\ktild=0.01,$ 0.1, and 0.2.
The scalar resonances are produced through gluon-gluon fusion,
and \RS graviton resonances through both gluon-gluon fusion and quark-antiquark annihilation.
In the \RS model,
the first mechanism accounts for approximately 90\% of the production cross section.

The SM background mostly arises from the direct production of two photons,
the production of $\gamma+$jets events in which jet fragments are misidentified as photons,
and the production of multijet events with misidentified jet fragments.
These backgrounds are simulated
with the \SHERPA\,2.1~\cite{Gleisberg:2008ta},
\MGAMC\,2.2~\cite{Alwall:2014hca}
(interfaced with \PYTHIA\,8.2 for parton showering and hadronization),
and \PYTHIA\,8.2 generators, respectively,
using the CT10NLO~\cite{Lai:2010vv},
NNPDF3.0~\cite{Ball:2014uwa},
and NNPDF2.3 PDF sets,
again respectively.
The \PYTHIA tune CUETP8M1~\cite{Khachatryan:2015pea} is used.

For both the signal and background samples,
the detector response is simulated
using the \GEANTfour package~\cite{Agostinelli:2002hh}.
The simulated samples incorporate additional {\Pp\Pp} interactions within the same or a
nearby
bunch crossing (pileup) and are weighted to reproduce the measured distribution of the
number of interactions per bunch crossing.  The average number of interactions per bunch
crossing is 18, with an RMS of 4.

\section{Event selection and diphoton mass spectrum}
\label{sec:event_sel}

The trigger requirements,
photon identification criteria,
and event selection procedures are
described in Ref.~\cite{cms-dipho-2015}.
Some details are given below.
Energy deposits in the ECAL compatible with the shower shape expected for
a photon are clustered together to define a photon candidate.
Variations in the crystal transparency during the data collection period are
corrected for using a dedicated monitoring system,
and the single-channel response is equalized based on collision
data~\cite{CMS:EGM-14-001}.
A multivariate regression technique~\cite{CMS:EGM-14-001} is used to correct the photon
energy for the incomplete containment of the shower in the clustered crystals, the shower
losses for photons that convert before reaching the calorimeter, and the effects of
pileup.
The interaction vertex is selected using the algorithm described in
Ref.~\cite{Khachatryan:2014ira},
which combines information on the correlation between
the diphoton system and the recoiling tracks,
the average transverse momentum (\pt) of the recoiling tracks,
and, when available,
directional information from reconstructed photon conversions.
For resonances with a mass above 500\GeV,
the fraction of events in which the interaction vertex is correctly
assigned is approximately 90\%.
For each photon candidate,
the transverse size of the electromagnetic cluster in the $\eta$ coordinate
must be compatible with that expected for a photon from a hard interaction,
and the ratio of the associated energy in the HCAL to the photon energy must
be less than 0.05.

Photon candidates are required to have $\pt>75\GeV$ and to be reconstructed
within $\abs{\eta}<2.5$.
Candidates in the transition region between the barrel and endcap detectors
($1.44<\abs{\eta}<1.57$),
where the reconstruction efficiency is not well described by the simulation,
are rejected.
Photon candidates associated with electron tracks that are incompatible with conversion
tracks are rejected~\cite{CMS:EGM-14-001}.
Photon candidates are required to be isolated.
There are two isolation criteria,
both of which are imposed:
i)~the sum of the scalar \pt of charged hadron candidates from the
interaction vertex that lie within a cone
of radius $R=\sqrt{\smash[b]{(\Delta\eta)^2+(\Delta\phi)^2}}=0.3$
around the photon candidate must be less than 5\GeV,
where charged hadron candidates identified as conversion
tracks associated with the photon candidate are excluded;
ii)~the sum of the scalar \pt of additional neutral electromagnetic candidates
within this same cone must be less than 2.5\GeV, after the contribution of additional
interactions in the same bunch crossing has been removed.

The identification and trigger efficiencies are measured
as functions of photon \pt
using data events containing a Z boson decaying
to a {\MM} pair in association with a photon,
or to an {\EE} pair where the electrons are treated as if they were photons~\cite{CMS:EGM-14-001}.
The efficiency of the photon selection procedure in the kinematic range considered in the
analysis is above 90\% and 85\% for candidates in the barrel and endcap regions,
respectively.
The ratio between the efficiencies measured in data and simulation is found to be lower
than 1 by 3.5\% for photons in the barrel region and by 6.5\% for photons in the endcap
region.
No significant \pt dependence of the efficiency ratios is observed, and a \pt-independent
correction is applied to the normalization of the simulated event samples to account for
this difference.

The photon candidates in an event are grouped into all possible pairs.
At least one photon candidate in the pair must have $\abs{\eta}<1.44$,
\ie, be reconstructed in the barrel.
Events with both photons in the endcaps are not considered, since their inclusion would
increase the signal efficiency by only a few percent, at the cost of introducing a large
background.
Photon pairs are divided into two categories.
The first category, denoted ``EBEB'',
contains pairs for which both candidates lie in the barrel.
For the second category, denoted ``EBEE'',
one candidate lies in the barrel and the other in an endcap.
The invariant mass \mgg of the pair must satisfy $\mgg>230\GeV$
for EBEB candidates and  $\mgg>330\GeV$ for EBEE candidates.
The fraction of events in which more than one photon pair satisfies
the selection criteria is approximately 1\%.
In these cases, only the pair with the largest
scalar sum of photon \pt is retained.

The selection efficiency times acceptance for signal events varies between 50\% and 70\%,
depending on the signal hypothesis.
Because of the different angular distribution of the decay products,
the kinematic acceptance for the \RS graviton resonances is lower
than that of scalar resonances.
For $\mX<1\TeV$ the difference is approximately 20\%.
The two acceptances are similar for $\mX > 3\TeV$.

The event selection procedure described above
is the same as the one documented in~\cite{cms-dipho-2015}.
It was finalized on the basis of studies with
simulated signal and background event samples prior to inspection of
the data in the search region of the diphoton invariant mass distribution,
which is defined as $\mgg > 500\GeV$.

A total of 6284 (2791) photon pairs are selected in the EBEB (EBEE) category.
Of these, 461 (800) pairs have an invariant mass above 500\GeV.
According to simulation, the direct production of two photons accounts, respectively, for
90\% and 80\% of the background events selected in the EBEB and EBEE categories.
This prediction is tested in data using the method described in Ref.~\cite{smdipho_7TeV}
and good agreement is found between data and simulation.

\begin{figure}[bthp]
    \centering
    \includegraphics[width=\cmsFigWidth]{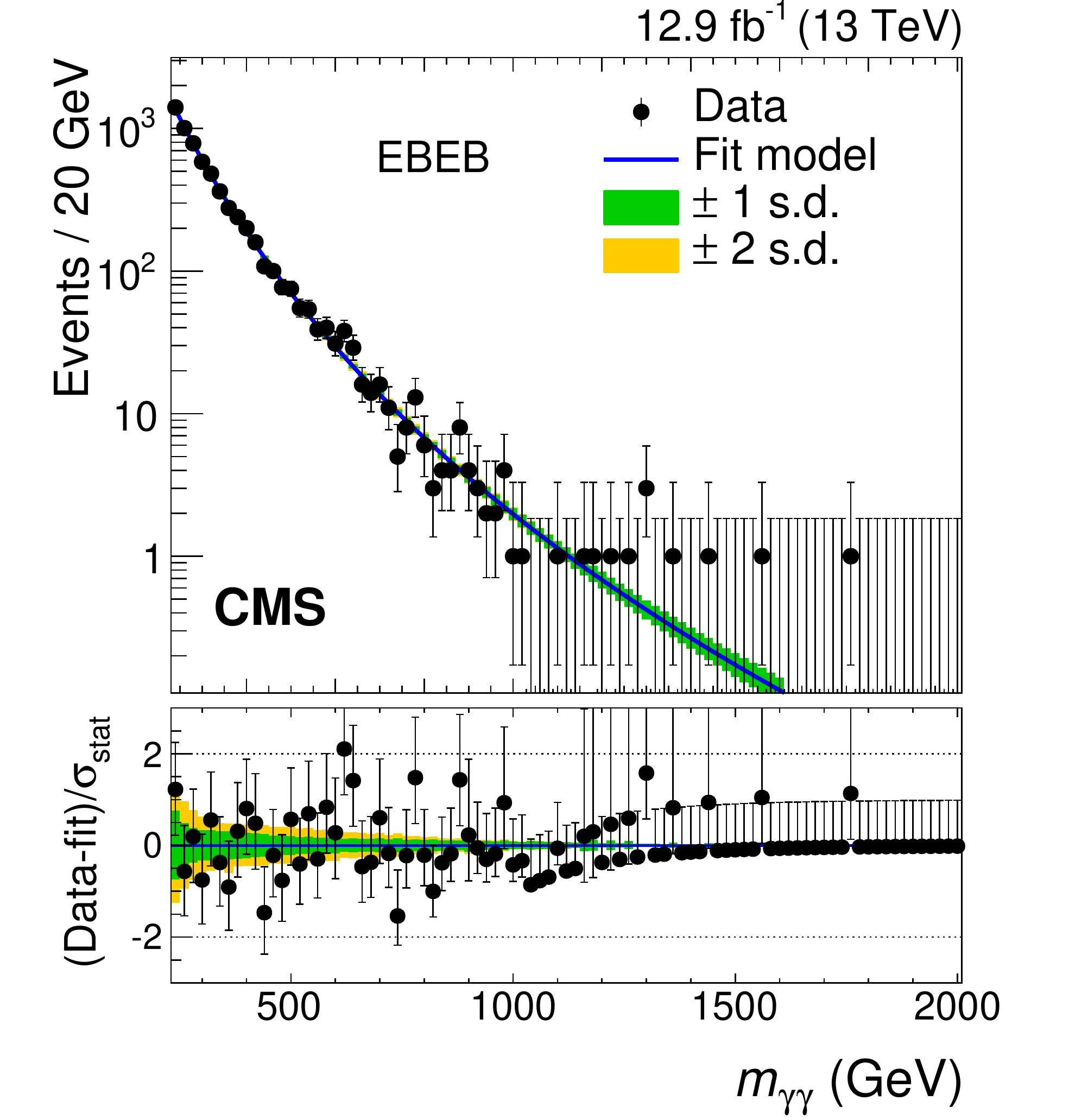}
    \includegraphics[width=\cmsFigWidth]{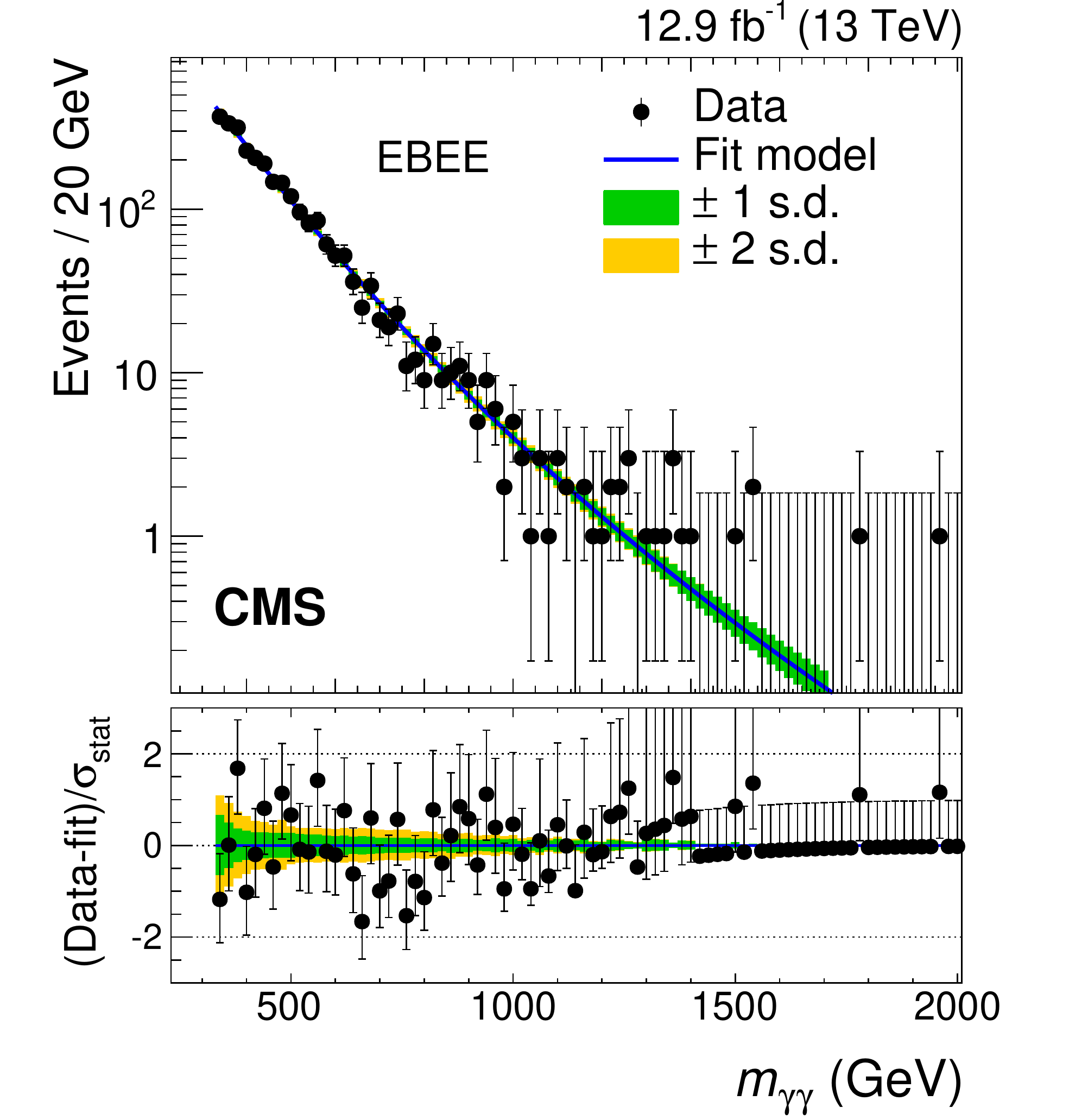}
    \caption{
      The observed invariant mass spectra \mgg for selected events in
      the (\cmsLeft) EBEB and (\cmsRight) EBEE categories.
      There are no selected events with $\mgg>2000\GeV$.
      The solid lines and the shaded bands show the results of likelihood fits to the data
      together with the associated 1 and 2 standard deviation statistical uncertainty bands.  The ratio
      of the difference between the data and the fit to the statistical uncertainty in the
      data is given in the lower plots.
      \label{fig:fits}
    }
\end{figure}

The diphoton invariant mass distribution of the selected events is shown in
Fig.~\ref{fig:fits}, for both the EBEB and EBEE categories.
We perform an independent maximum likelihood fit to the data
in each category using the function
\begin{equation}
  f(\mgg) = \mgg^{a + b \, \log( \mgg )}
  \label{eq:mgg-sm}.
\end{equation}
This parametric form is chosen to model the background
in the hypothesis tests discussed below.
The results of the fits are shown in Fig.~\ref{fig:fits}.

\section{Likelihood fit}

A simultaneous fit to the invariant mass spectra of events
in the EBEB and EBEE event categories is performed
to determine the compatibility of the data with the
background-only and the signal+background hypotheses.
The test statistic is based on the profile likelihood ratio:
\begin{equation}
q(\mu) = -2 \log \frac{ L(\mu S + B | \hat{\vec{\theta}}_{\mu} ) }
    {L (\hat\mu S + B | \hat{\vec{\theta}} )},
\end{equation}
where $S$ and $B$ represent the probability density functions
for resonant diphoton production and for the SM background,
respectively.
The parameter $\mu$ is the so-called signal strength,
while $\vec{\theta}$ represents the nuisance parameters of the model,
used to account for systematic uncertainties.
The $\hat{\theta}$ notation indicates the best fit value of the parameter~$\theta$ for any $\mu$ value,
while $\hat{\theta}_\mu$ denotes the best fit value of~$\theta$ for a fixed value~$\mu$.

To set upper limits on the rate of resonant diphoton production,
the modified frequentist method known as \CLs~\cite{Junk1999,bib-cls} is used,
following the prescription described in Ref.~\cite{ATLAS:1379837}.
The compatibility of the observation with the background-only hypothesis is evaluated
by computing the background-only $p$-value.
The latter is defined as the probability,
in the background-only hypothesis,
for $q(0)$  to exceed the value observed in data.
This quantity, the ``local $p$-value'' $p_0$,
does not take into account the fact that many signal hypotheses are tested.

Asymptotic formulas~\cite{Cowan:2010js} are used in the
calculations of exclusion limits and local $p$-values.
The accuracy of the asymptotic approximation in the estimation of
exclusion limits and significance is
studied, using pseudo-experiments, for a subset of the hypothesis tests and is found to be
about 10\%.

The signal shape in \mgg is determined from the convolution of the intrinsic shape
of the resonance and the CMS detector response to photons.
The intrinsic shape is taken from the \PYTHIA 8.2 generator.
A grid of mass points with 125\GeV spacing, in the range 500--4500\GeV, is used.
The resulting shapes are interpolated to intermediate points
using a parametric description of the distribution.
The detector response is determined using fully simulated
signal samples of small intrinsic width,
corrected through Gaussian smearing to agree with measurements
based on $\PZ\to\EE$ data.
Nine uniformly spaced mass hypotheses in the range 500--4500\GeV are employed.
The signal mass resolution,
quantified through the ratio of the
full width at half maximum of the distribution,
divided by 2.35,
to the peak position,
is approximately 1.0\% and 1.5\% for the EBEB and EBEE categories, respectively.
The signal normalization coefficients are proportional to the product of the kinematic
acceptance and the signal efficiency within the acceptance region. These are computed, for
each category, in simulated samples and interpolated to intermediate points using
quadratic functions of \mX and \GammaOm.

The background shape in \mgg is described by the
parametric function given by Eq.~(\ref{eq:mgg-sm}).
The values of the parameters $a$ and $b$ are determined by the fit to data,
with separate values for the EBEB and EBEE categories,
and are treated as unconstrained nuisance parameters in the hypothesis tests.

The accuracy of the background parameterization is assessed using
simulation and is quantified by studying the difference between the true and
predicted numbers of background events in several \mgg intervals in the search region.
The relative widths of the intervals,
defined by $2(x_1-x_2)/(x_1+x_2)$ with $x_1$ and $x_2$ the lower and upper bin edges,
range between 2\% and 15\%.
Pseudo-experiments are drawn from the mass spectrum predicted
by the simulation and are fit with the chosen background model.
The total number of events in each pseudo-experiment is taken from a Poisson
distribution whose mean is set equal to the observation in data.
For each interval,
the distribution of the pull variable,
defined as the difference between the true and predicted numbers of
events divided by the estimated statistical uncertainty,
is constructed.
If the absolute value of the median of this distribution is
found to be above 0.5 in an interval,
an additional uncertainty is assigned to the background parametrization.
A modified pull distribution is then constructed,
increasing the statistical uncertainty in the fit by an extra term,
denoted the ``bias term''.
The bias term is parametrized as a smooth function of \mgg,
which is tuned in such a manner that the absolute value of the median of the
modified pull distribution is less than 0.5 in all intervals.
The amplitude of the bias term function is comparable to that of the 1 standard deviation
bands in Fig.~\ref{fig:fits}.
This additional uncertainty is included in the likelihood function by adding
to the background model a component having the same shape as the signal.
The normalization coefficient of this component is constrained to have a Gaussian
distribution of mean zero,
with a width equal to the integral of the bias term
function over the full width at half maximum of the tested signal shape.
The inclusion of this additional component has the effect of avoiding
falsely positive or falsely negative tests that could be induced by a
mismodeling of the background shape, and it reduces the sensitivity of the
analysis by at most~10\%.

\section{Systematic uncertainties}

The impact of systematic uncertainties in this analysis
is smaller than that of the statistical uncertainties.
The parametric background model has no associated systematic uncertainties
except for the bias term uncertainty described in the previous section.
Since the background shape coefficients $a$ and $b$
[Eq.~(\ref{eq:mgg-sm})]
are treated as unconstrained nuisance parameters,
the associated uncertainties are statistical.

The systematic uncertainties in the signal normalization
associated with the integrated luminosity,
the selection efficiency,
and the PDFs
are 6.2\%, 6.0\%, and 6.0\%, respectively.
The uncertainty in the integrated luminosity is estimated from
beam scans performed in August 2016,
utilizing the methods of Ref.~\cite{CMS-PAS-LUM-15-001}.
The uncertainty associated with the PDFs is evaluated
by comparing the overall selection efficiency obtained with
the CT10~\cite{Lai:2010vv}, MSTW08~\cite{Martin2009}, and NNPDF2.3~\cite{Ball:2012cx} PDF
sets and taking the largest deviation over all tested signal hypotheses.
A 1\% uncertainty is associated with the level of knowledge of the energy scale
and accounts for the uncertainty in the energy scale at the {\PZ} boson
peak and its extrapolation to higher masses.
A 10\% uncertainty is assigned to the knowledge of the photon energy resolution,
corresponding to the uncertainty in the estimated additional Gaussian smearing determined
at the {\PZ} boson peak.

\section{Results for the 2016 data}
\label{sec:results:13fb}

\begin{figure}[!htb]
    \centering
    \includegraphics[width=\cmsFigMedWidth]{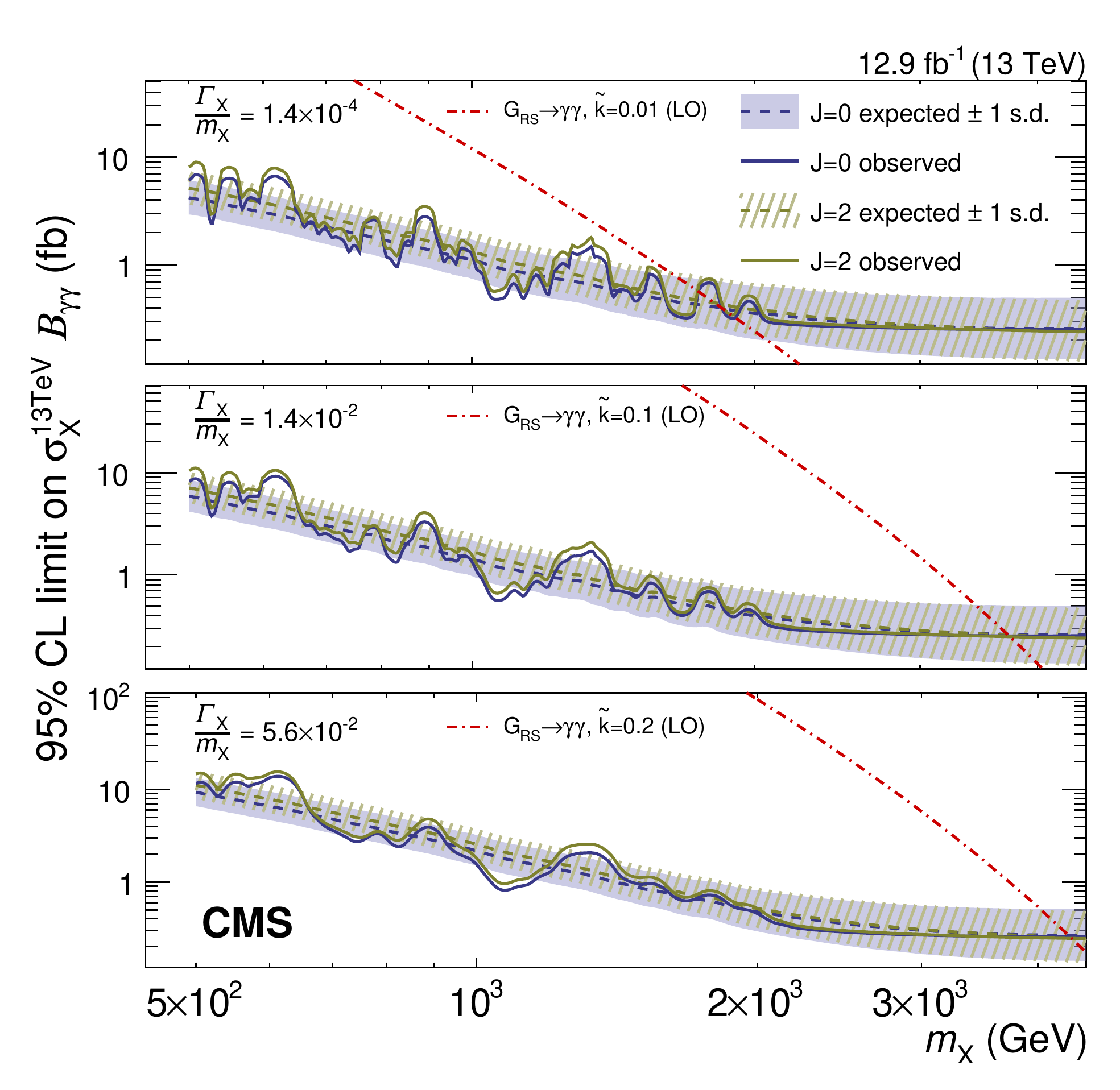}
    \caption{
      The 95\% \CL upper limits on the production of diphoton resonances as a
      function of the resonance mass \mX, from the analysis of data collected in 2016.
      Exclusion limits for the scalar and \RS graviton signals are given by the grey (darker) and
      green (lighter) curves, respectively.  The observed limits are shown by the solid lines,
      while the median expected limits are given by the dashed lines together with their
      associated 1 standard deviation uncertainty bands.  The leading-order production cross
      section for diphoton resonances in the \RS graviton model is shown for three values of the
      dimensionless coupling parameter \ktild together with the exclusion upper limits
      calculated for the corresponding three values of the width relative to the mass, \GammaOm.
      Shown are the results for (upper) a narrow width, (middle) an intermediate-width, and
      (lower) a broad resonance.
      \label{fig:limits}
    }
\end{figure}

\begin{figure}[!htb]
    \centering
    \includegraphics[width=\cmsFigWidth]{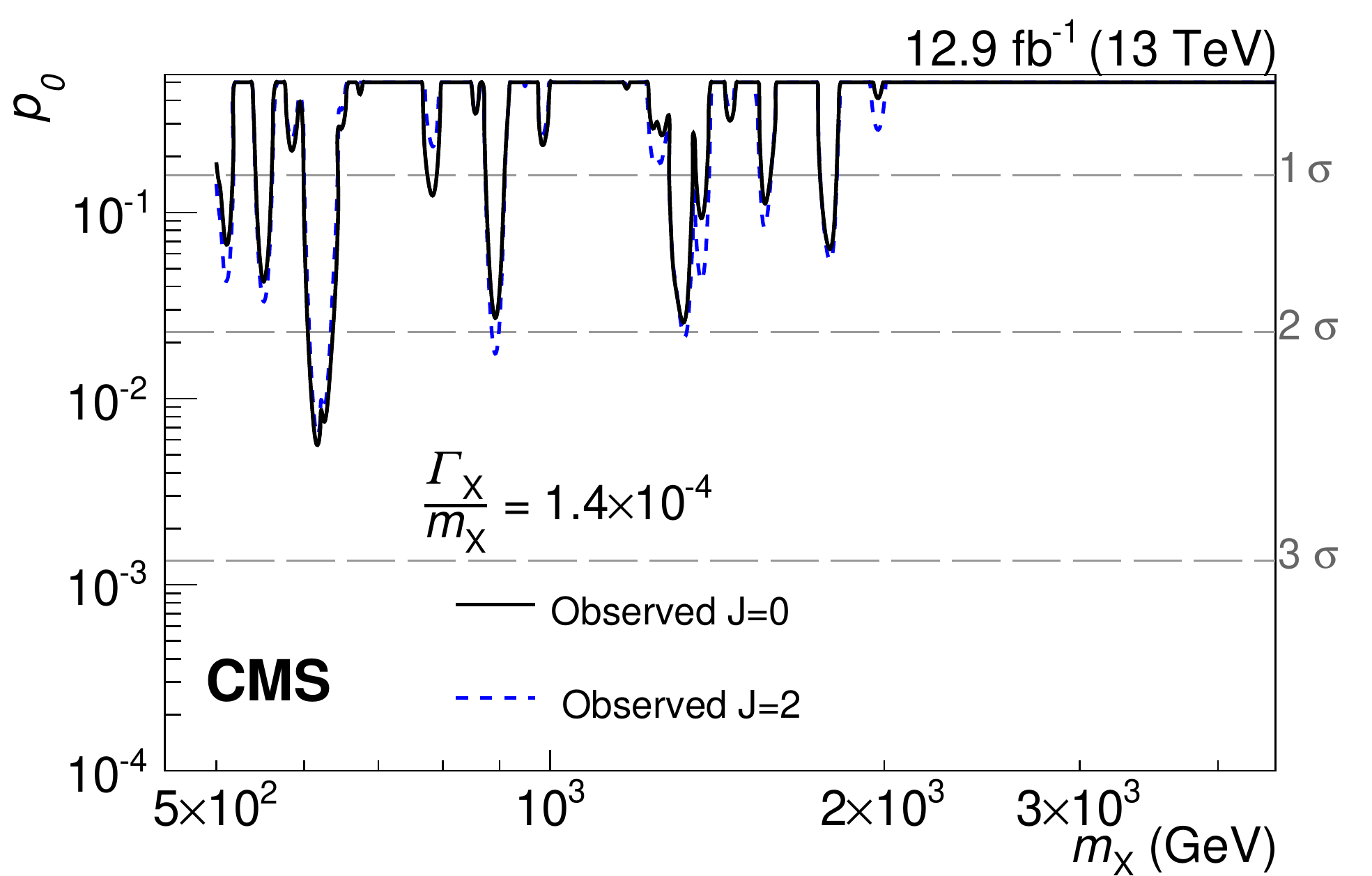}
    \includegraphics[width=\cmsFigWidth]{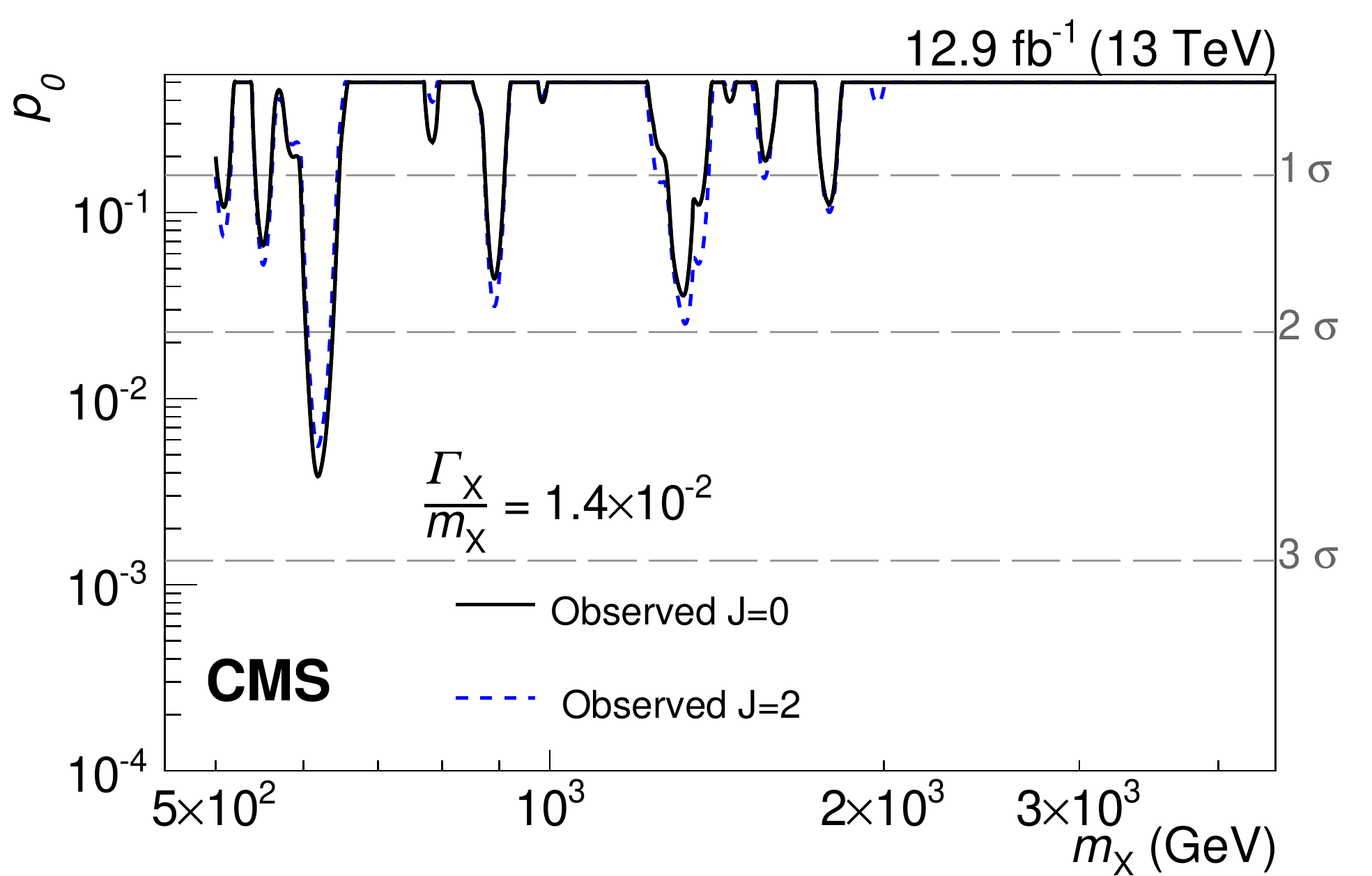}  \\
    \includegraphics[width=\cmsFigWidth]{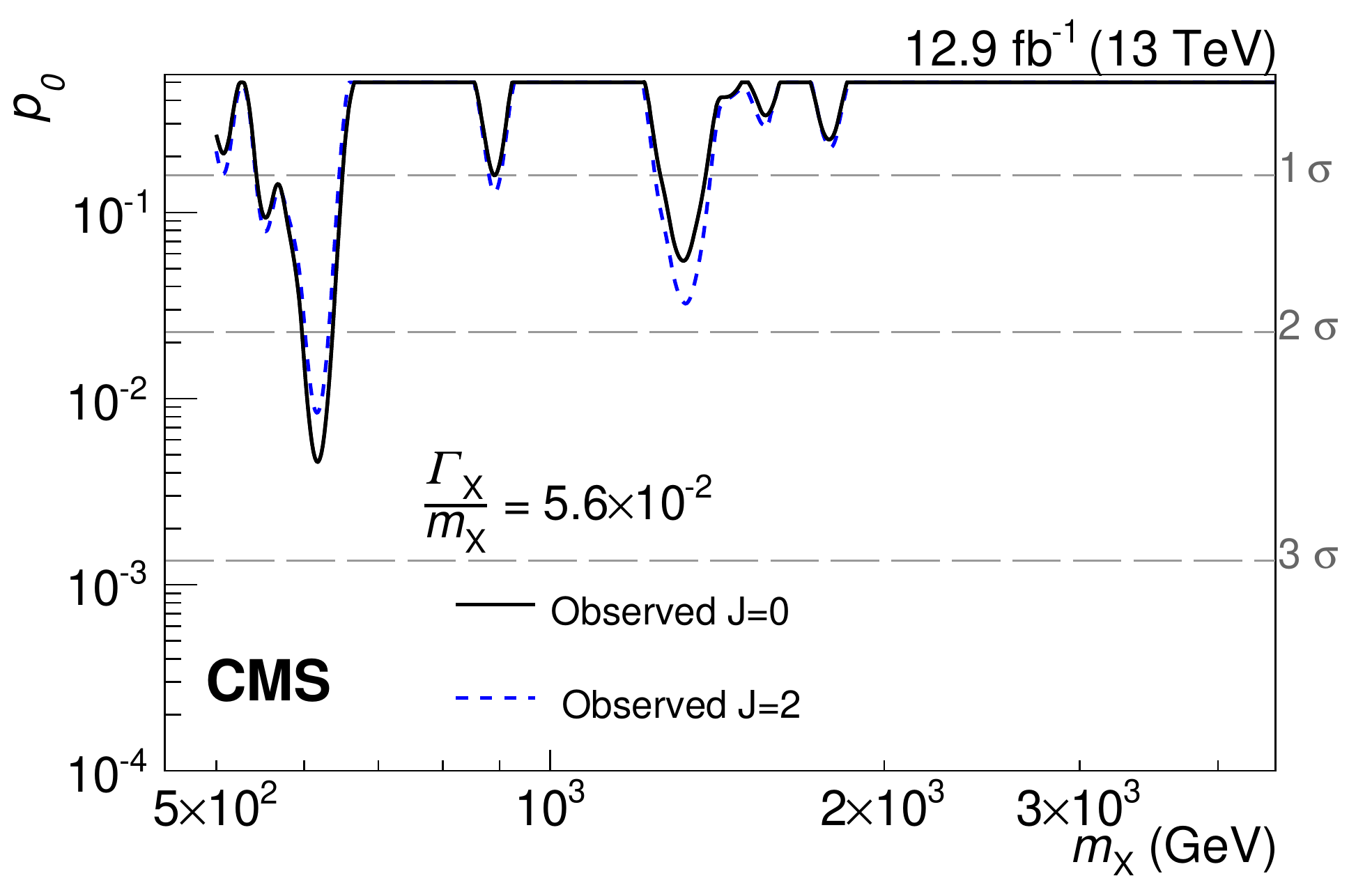}
    \caption{
      Observed background-only $p$-values for resonances with
      (\cmsULeft) $\GammaOm = 1.4\times 10^{-4}$,
      (\cmsURight) $1.4\times 10^{-2}$,
      and (bottom) $5.6\times 10^{-2}$ as a function of the resonance mass \mX,
      from the analysis of data collected in 2016.
      The solid black and dashed blue lines correspond to spin-0 and spin-2 resonances,
      respectively.
      \label{fig:pvalues}
    }
\end{figure}

The observed and expected 95\% confidence level (CL)
upper limits on the product of the production cross section
($\sigma_\mathrm{X}^{13\TeV}$) and branching fraction
to two photons ($\mathcal{B}_{\gaga}$) for
scalar and \RS graviton resonances are shown in Fig.~\ref{fig:limits}.
Using the LO cross sections from \PYTHIA\,8.2,
\RS gravitons with masses below 1.75, 3.75, and 4.35\TeV
are excluded for $\ktild=0.01$, 0.1, and 0.2, respectively,
corresponding to $\GammaOm=1.4\times10^{-4}$, $1.4\times10^{-2}$, and $5.6\times10^{-2}$.

The value of $p_{0}$ for different signal hypotheses is shown in
Fig.~\ref{fig:pvalues}.
The largest excess is observed for $\mX\approx 620\GeV$,
and has a local significance of approximately 2.4 and 2.7 standard deviations
for narrow spin-0 and \RS graviton signal hypotheses,
respectively.
After taking into account the effect of searching for several signal hypotheses,
i.e., searching over a range of widths and masses,
the significance of the excess is reduced to less than one standard deviation.
No excess is observed in the proximity of $\mX=750\GeV$.

\section{Combination with the 2012 and 2015 data}

The results obtained for the 2016 data are combined statistically
with those obtained for the data discussed in Ref.~\cite{cms-dipho-2015},
namely 19.7\fbinv of proton-proton collisions
recorded at $\sqrts=8\TeV$ in 2012~\cite{CMS-dipho-8TeV}
and 3.3\fbinv recorded at $\sqrts=13\TeV$ in 2015.
For a portion of the 2015 data (0.6\fbinv),
the CMS magnet was off (\boff),
while for the rest of the 2015 data and for all of
the 2012 and 2016 data,
the magnet was at its operational field strength (\bon).
The analysis of the \boff data from 2015 is
described in Ref.~\cite{cms-dipho-2015}.

The procedure followed for the combined analysis of 8 and 13\TeV data is the same as in
Ref.~\cite{cms-dipho-2015}.
The ratio of the 8 to the 13\TeV production cross section is computed
using $\PYTHIA\,8.2$,
for the two types of signal hypotheses considered:
scalar resonances and \RS graviton resonances.
The cross section ratio decreases from 0.27 and 0.29
at $\mX=500\GeV$ to 0.03 and 0.04 at $\mX=4\TeV$, for
the scalar and RS graviton resonance hypotheses, respectively.

Exclusion limits are set on the 13\TeV production cross section for both models,
and background-only $p$-values are computed for the signal hypotheses.

The correlation model between the systematic uncertainties associated with 8 and 13\TeV
data is described in Ref.~\cite{cms-dipho-2015}.
It assumes all uncertainties to be uncorrelated
except for those related to the knowledge of the
PDFs, which are taken to be fully correlated, and
those related to the knowledge of the photon energy scale,
which are taken to have a linear correlation of 0.5.
Taking the value of the linear correlation to be 0 or 1 would lead to
negligible changes in the results.
For the combination of the two 13~TeV data sets,
the background shape and the associated bias term uncertainties
are assumed to be fully correlated between the corresponding categories
of the 2015 (\bon) and 2016 data.
Independent background normalization coefficients are used for the two data sets.
The uncertainty in the signal selection efficiency is taken to be uncorrelated between the
2015 and 2016 data, to account for the large statistical contribution and for the effect on
the systematic contribution arising from changes in the data taking conditions,
particularly in the instantaneous luminosity.
The uncertainty in the knowledge of the integrated luminosity
is treated as follows:
a 2.3\% uncertainty,
corresponding to the knowledge of the absolute luminosity scale
calibration determined with beam scans,
is taken to be fully correlated between
the 2015 (\bon) and 2016 data,
and additional uncertainties of 1.5\% and 5.8\%,
corresponding to the uncertainty in extrapolating
the scale calibration to the data collection conditions,
are applied, again respectively.
Finally, the photon energy scale uncertainties are taken to be fully correlated between
the two data sets, being dominated by the extrapolation to high energy.

\begin{figure}[!htb]
    \centering
    \includegraphics[width=\cmsFigMedWidth]{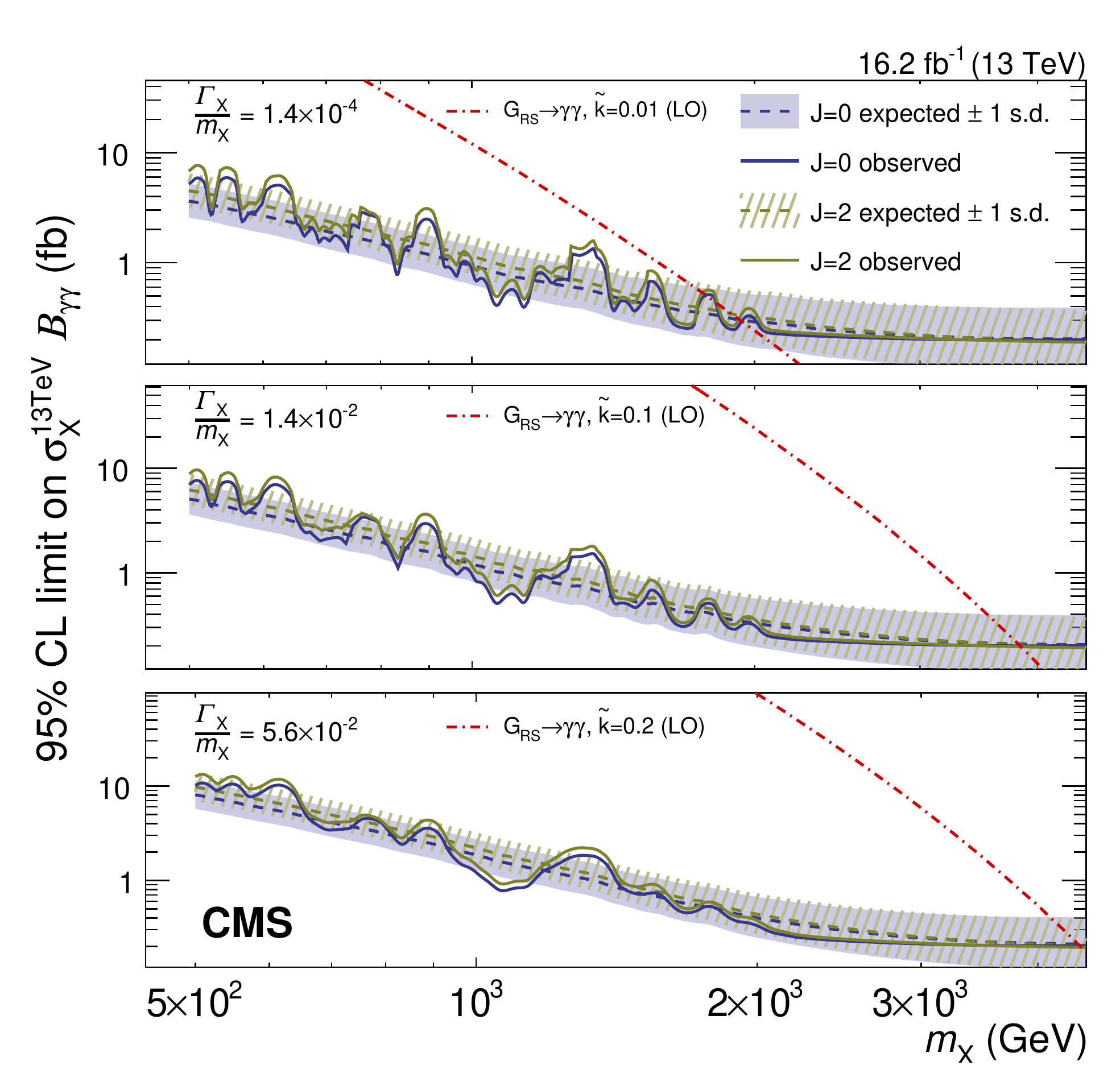}
    \caption{
     The 95\% \CL upper limits on the production of diphoton resonances as a
     function of the resonance mass \mX, from the combined analysis of data collected
     in 2015 and in 2016.
     Exclusion limits for the scalar and \RS graviton signals are given by the grey (darker) and
     green (lighter) curves, respectively.  The observed limits are shown by the solid lines,
     while the median expected limits are given by the dashed lines together with their
     associated 1 standard deviation uncertainty bands.  The leading-order production cross
     section for diphoton resonances in the \RS graviton model is shown for three values of the
     dimensionless coupling parameter \ktild together with the exclusion upper limits
     calculated for the corresponding three values of the width relative to the mass, \GammaOm.
     Shown are the results for (upper) a narrow width, (middle) an intermediate-width, and
     (lower) a broad resonance.
     \label{fig:limits:13TeV}
   }
\end{figure}

Figure~\ref{fig:limits:13TeV} shows the observed and expected 95\% CL upper limits
on the 13\TeV production cross section of the different signal hypotheses
obtained with the combined analysis of the 13\TeV data recorded in 2015 and 2016.
The upper limits on the production of scalar resonances decaying to two photons range from
about 10 to 0.2\fb, for resonance masses between 0.5 and 4.5\TeV.
Compared to the 2016 data alone,
the sensitivity is improved by approximately 10\% and 20\% at the high and low end
of the \mX search region, respectively.
Using the LO cross sections from \PYTHIA\,8.2,
\RS gravitons with masses below  3.85 and 4.45\TeV are excluded
for $\ktild=0.1$ and $0.2$, respectively.
For $\ktild=0.01$,
graviton masses below 1.95\TeV are excluded,
except for the region between 1.75 and 1.85\TeV.

\begin{figure*}[htb]
    \centering
    \includegraphics[width=\cmsFigWidth]{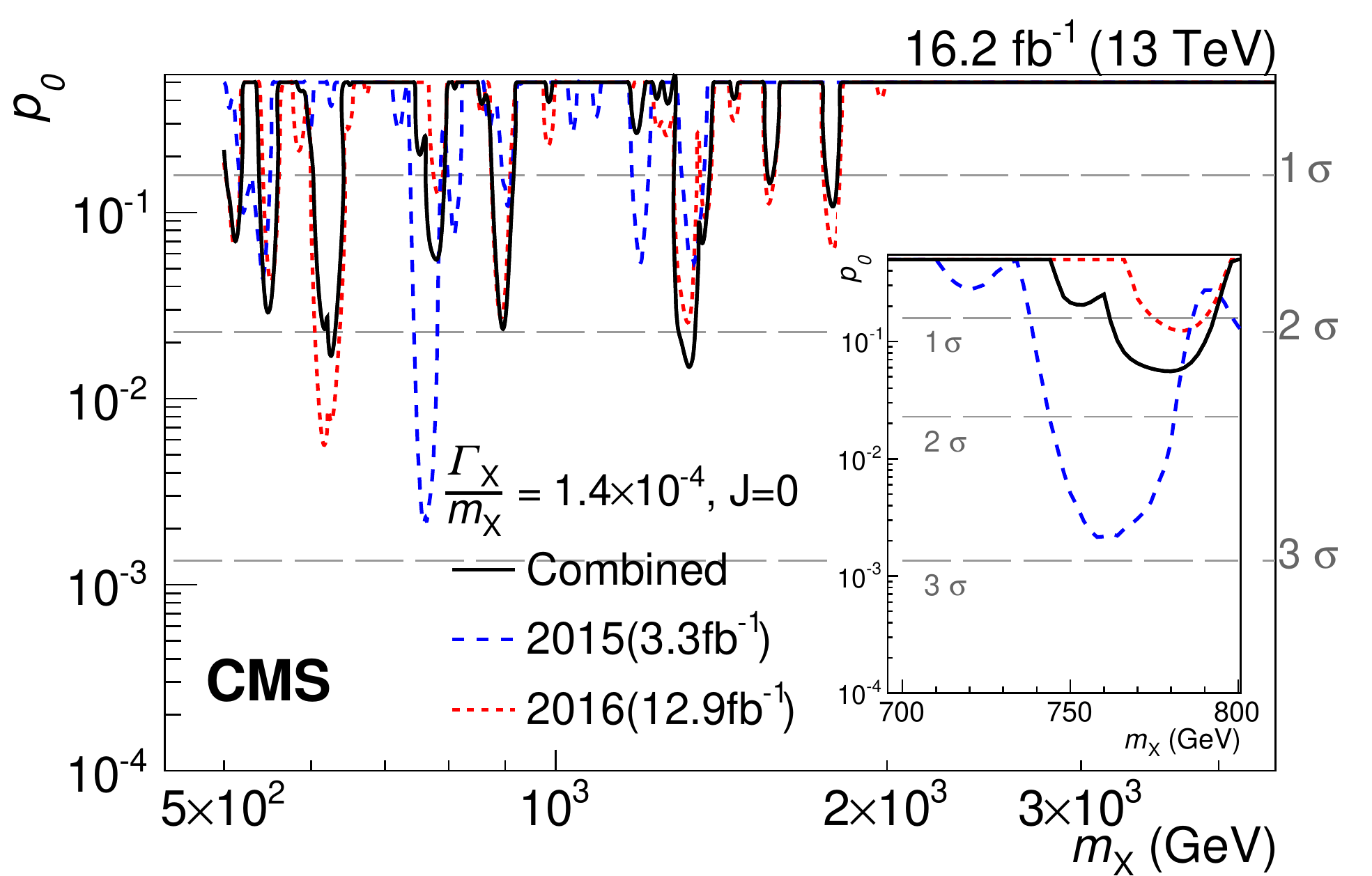}
    \includegraphics[width=\cmsFigWidth]{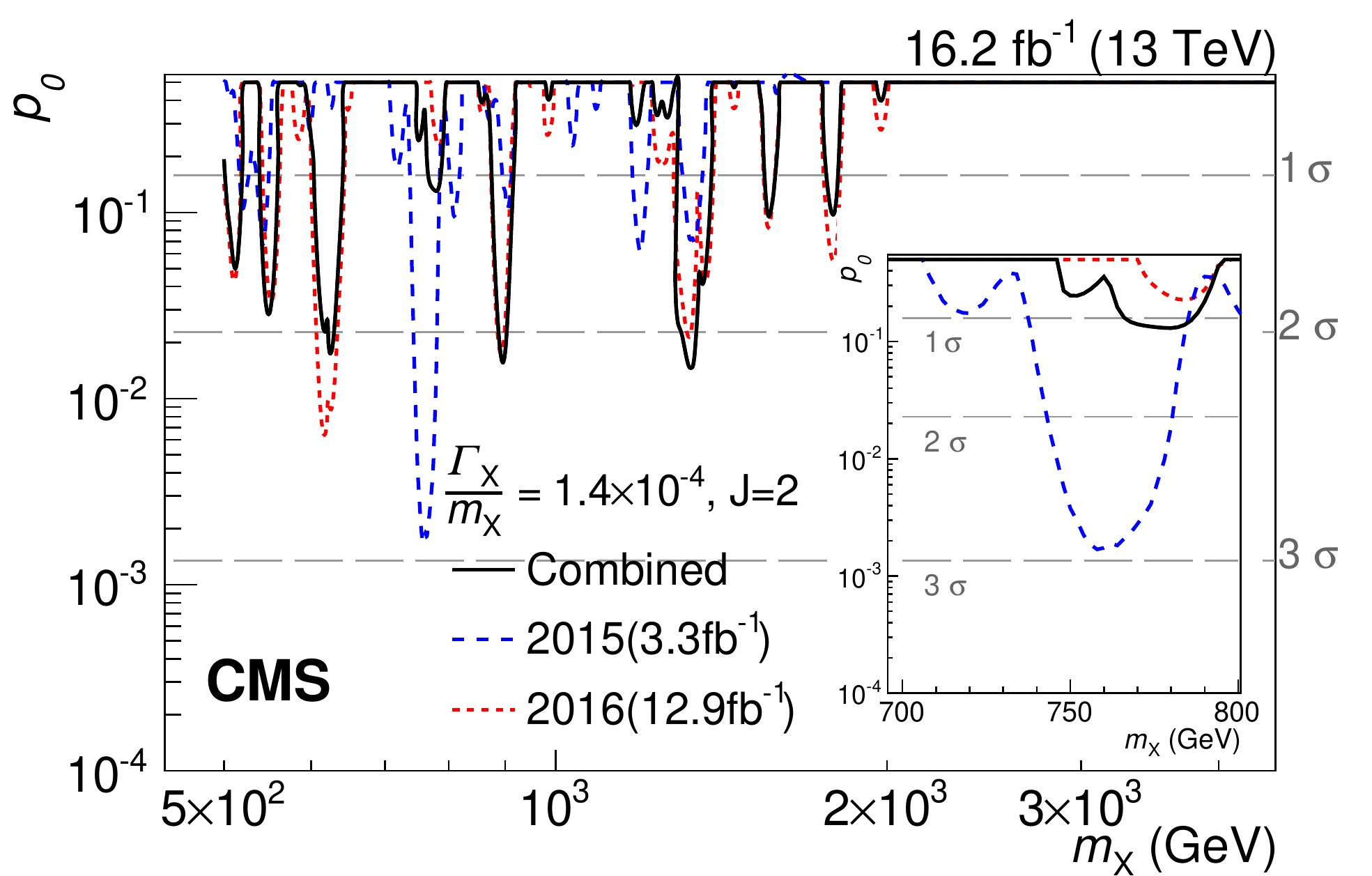}  \\[0.5ex]
    \includegraphics[width=\cmsFigWidth]{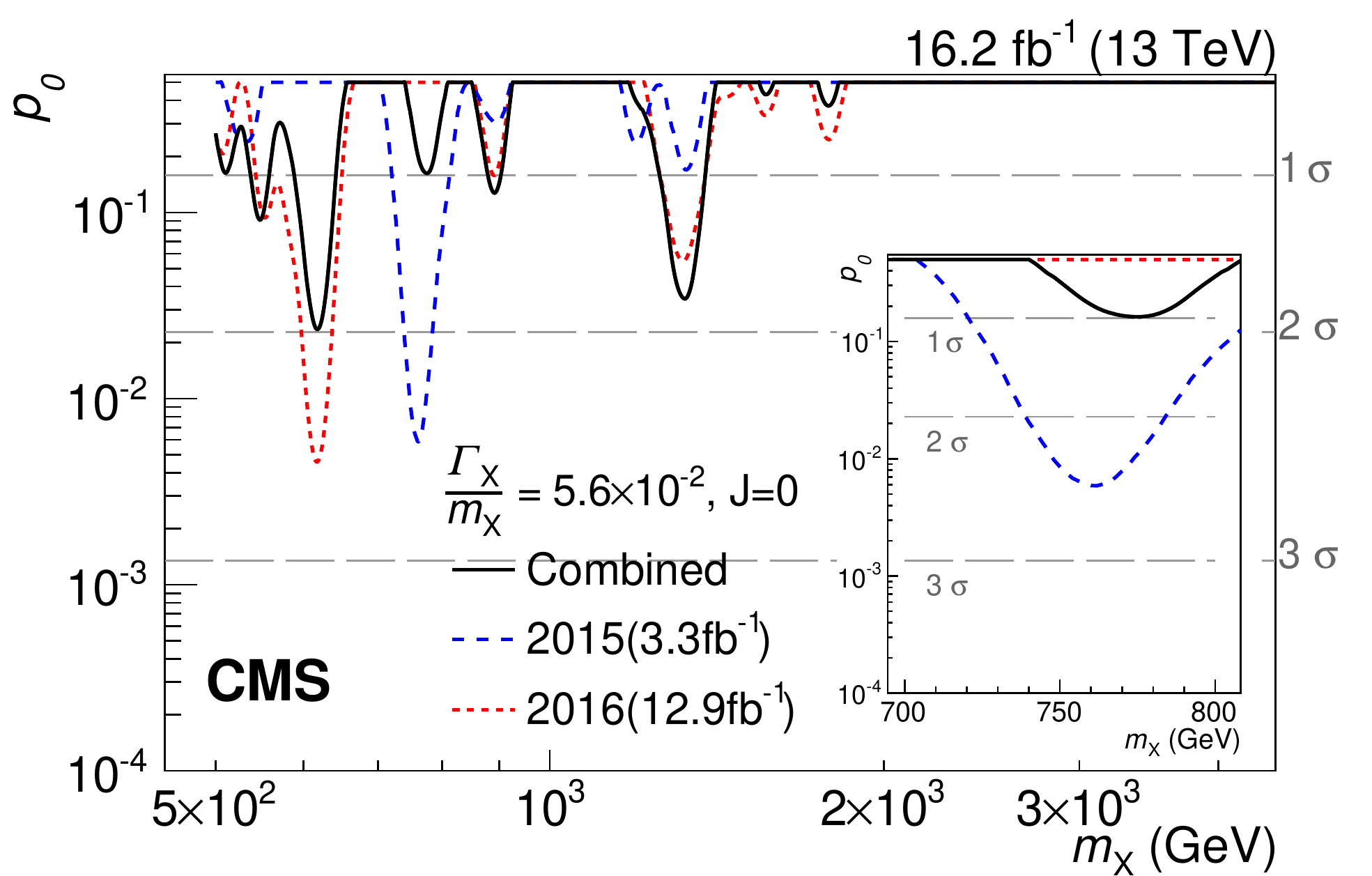}
    \includegraphics[width=\cmsFigWidth]{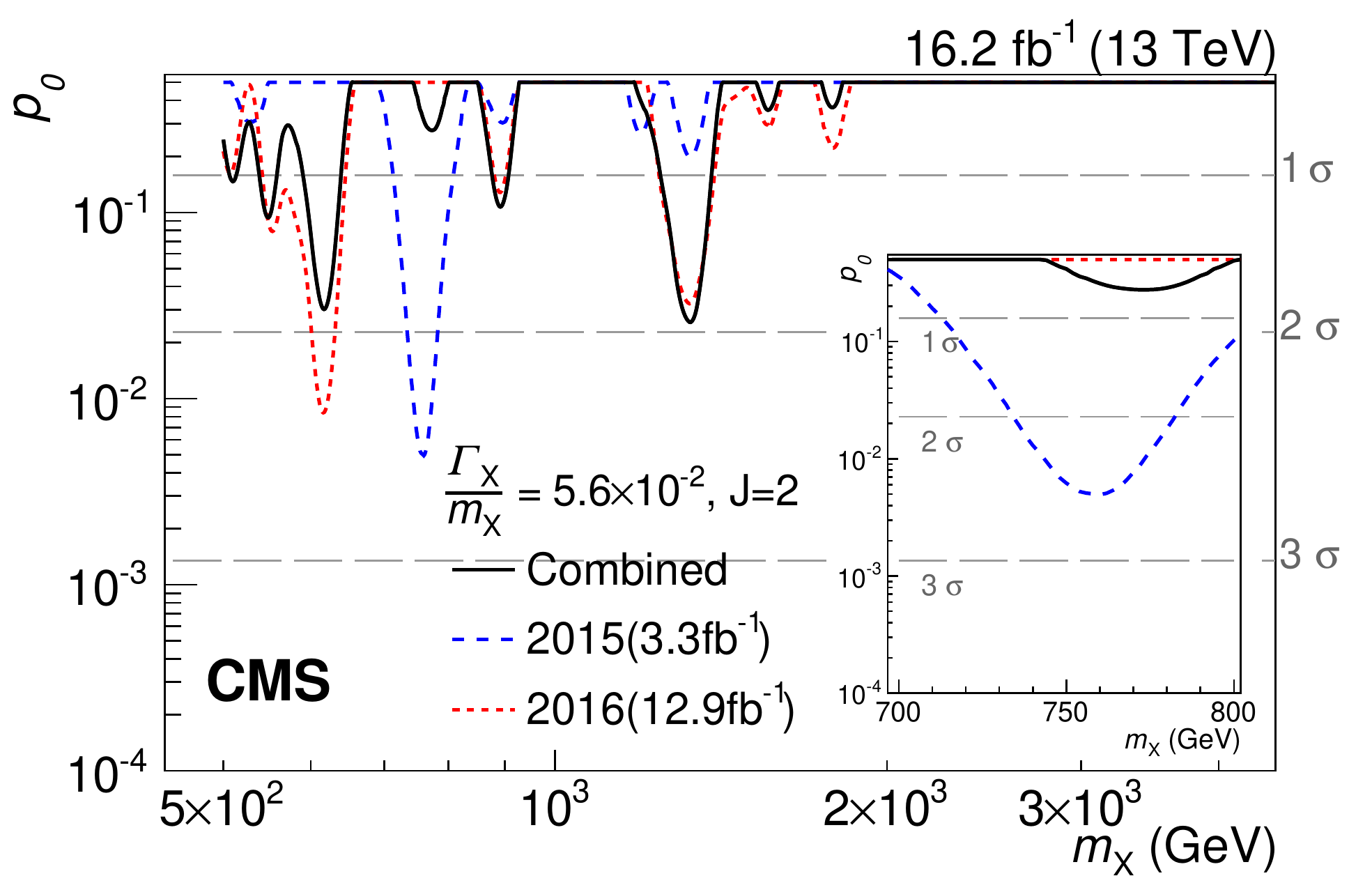}
    \caption{
      Observed background-only $p$-values for resonances with
      (upper) $\GammaOm = 1.4\times 10^{-4}$ and
      (lower) $5.6\times 10^{-2}$ as a function of the resonance mass \mX, from the
      combined analysis of data recorded in 2015 and 2016.
      The results obtained for the two individual data sets are also shown.
      The curves corresponding to the scalar and \RS graviton hypotheses are shown
      in left and right columns, respectively.
      The insets show an expanded region around $\mX=750\GeV$.
      \label{fig:pvalues:13TeV}
    }
\end{figure*}

The observed $p_0$ for $\GammaOm = 1.4\times 10^{-4}$ and $5.6\times 10^{-2}$ obtained
with the combined analysis of the 2015 and 2016 data is shown in
Fig.~\ref{fig:pvalues:13TeV}.
The largest excess is observed for $\mX \approx 1.3\TeV$ and
has a local significance of about 2.2 standard deviations,
corresponding to less than 1 standard deviation after accounting
for the effect of searching for several signal hypotheses.
For $\mX = 750\GeV$,
the 2.9 standard deviation local significance excess observed in
the 2015 data is reduced to 0.8 standard deviations.

\begin{figure}[htb]
    \centering
    \includegraphics[width=\cmsFigMedWidth]{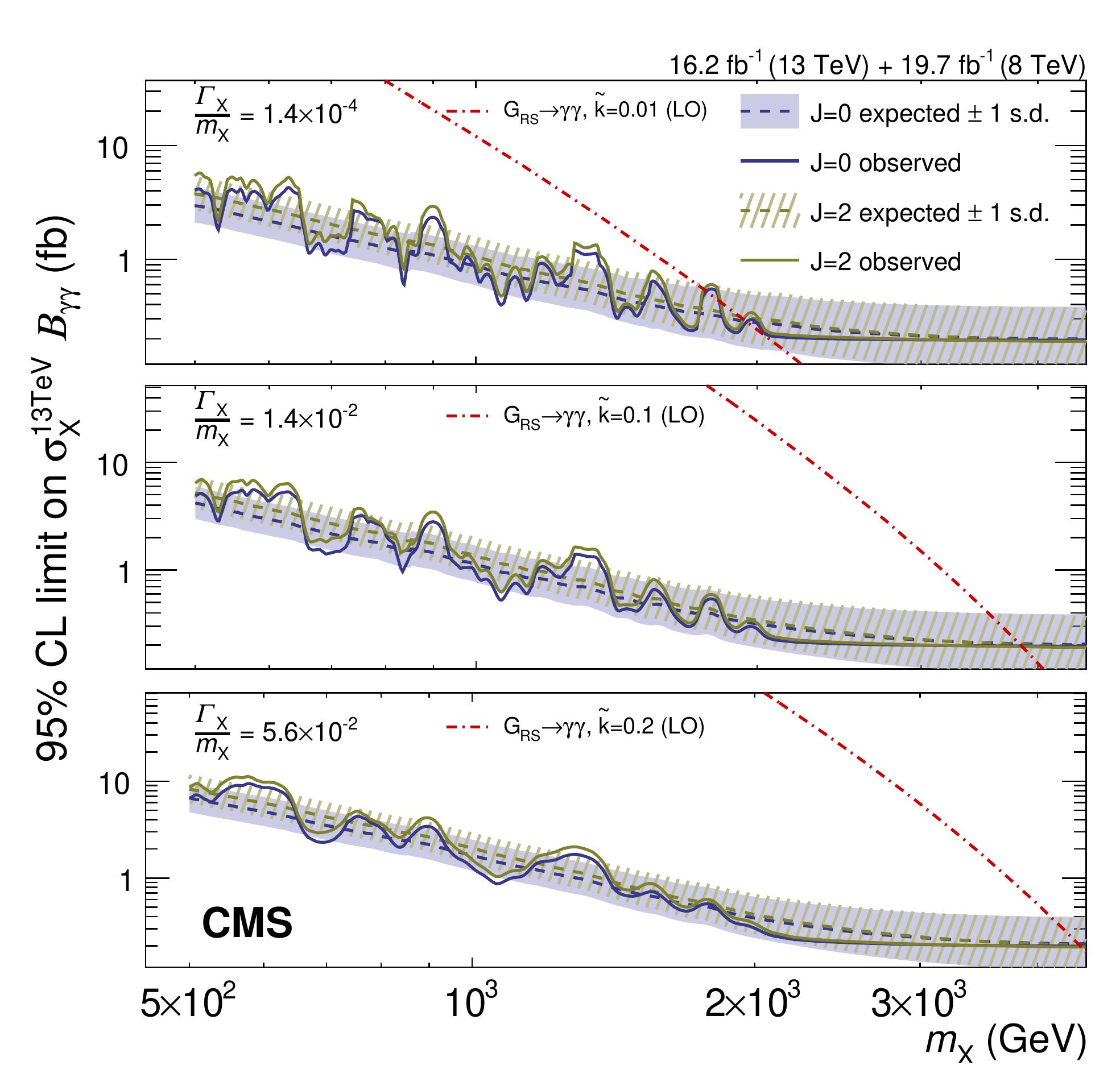}
    \caption{
      The 95\% \CL upper limits on the production of diphoton resonances as a
      function of the resonance mass \mX, from the combined analysis of the 8 and 13\TeV
      data.
      The 8\TeV results are scaled by the ratio of the 8 to 13\TeV cross sections.
      Exclusion limits for the scalar and \RS graviton signals are given by the grey (darker) and
      green (lighter) curves, respectively.  The observed limits are shown by the solid lines,
      while the median expected limits are given by the dashed lines together with their
      associated 1 standard deviation uncertainty bands.  The leading-order production cross
      section for diphoton resonances in the \RS graviton model is shown for three values of the
      dimensionless coupling parameter \ktild together with the exclusion upper limits
      calculated for the corresponding three values of the width relative to the mass, \GammaOm.
      Shown are the results for (upper) a narrow width, (middle) an intermediate-width, and
      (lower) a broad resonance.
      \label{fig:limits:13TeV_8TeV}
    }
\end{figure}

\begin{figure*}[htb]
    \centering
    \includegraphics[width=\cmsFigWidth]{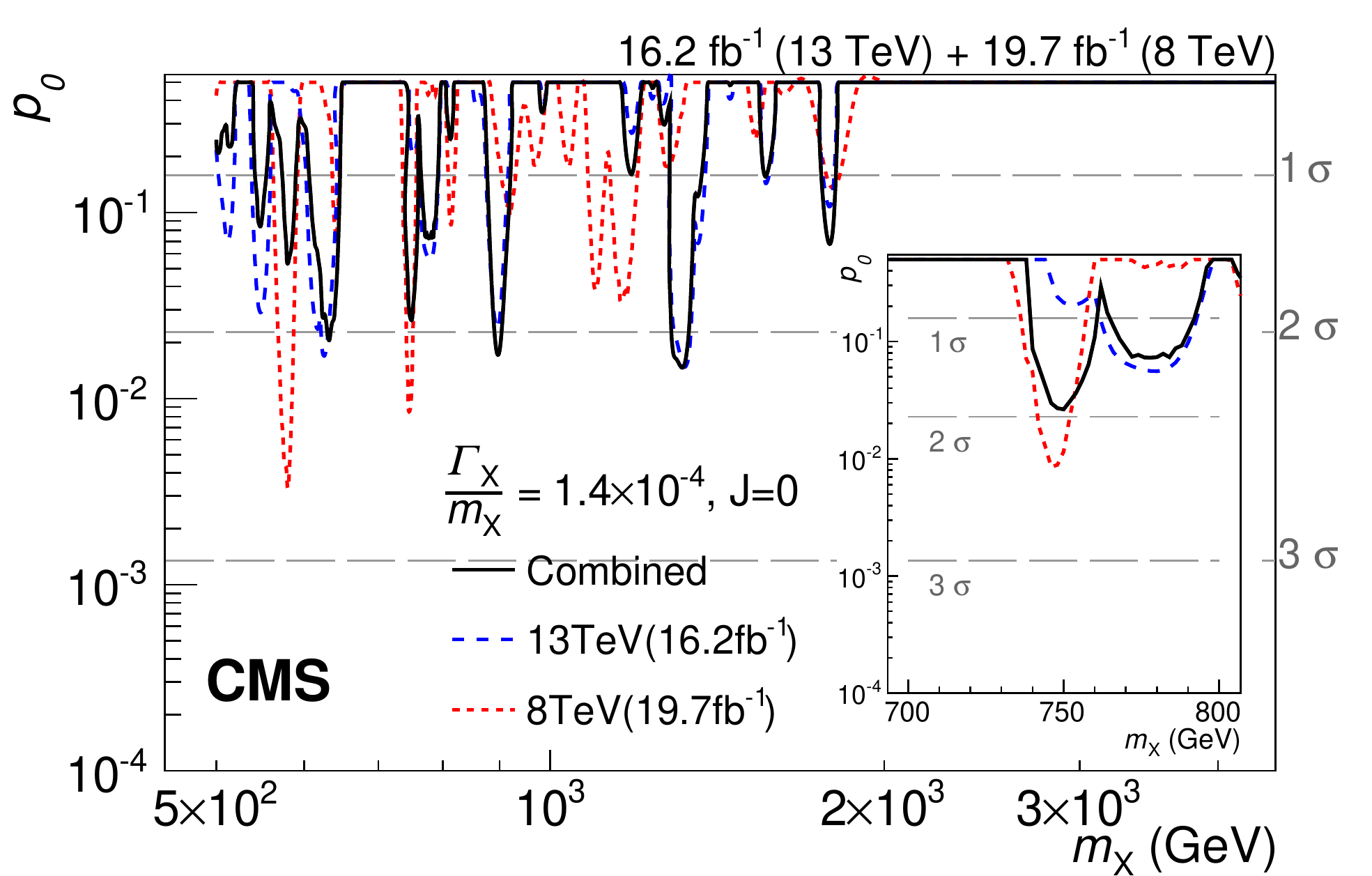}
    \includegraphics[width=\cmsFigWidth]{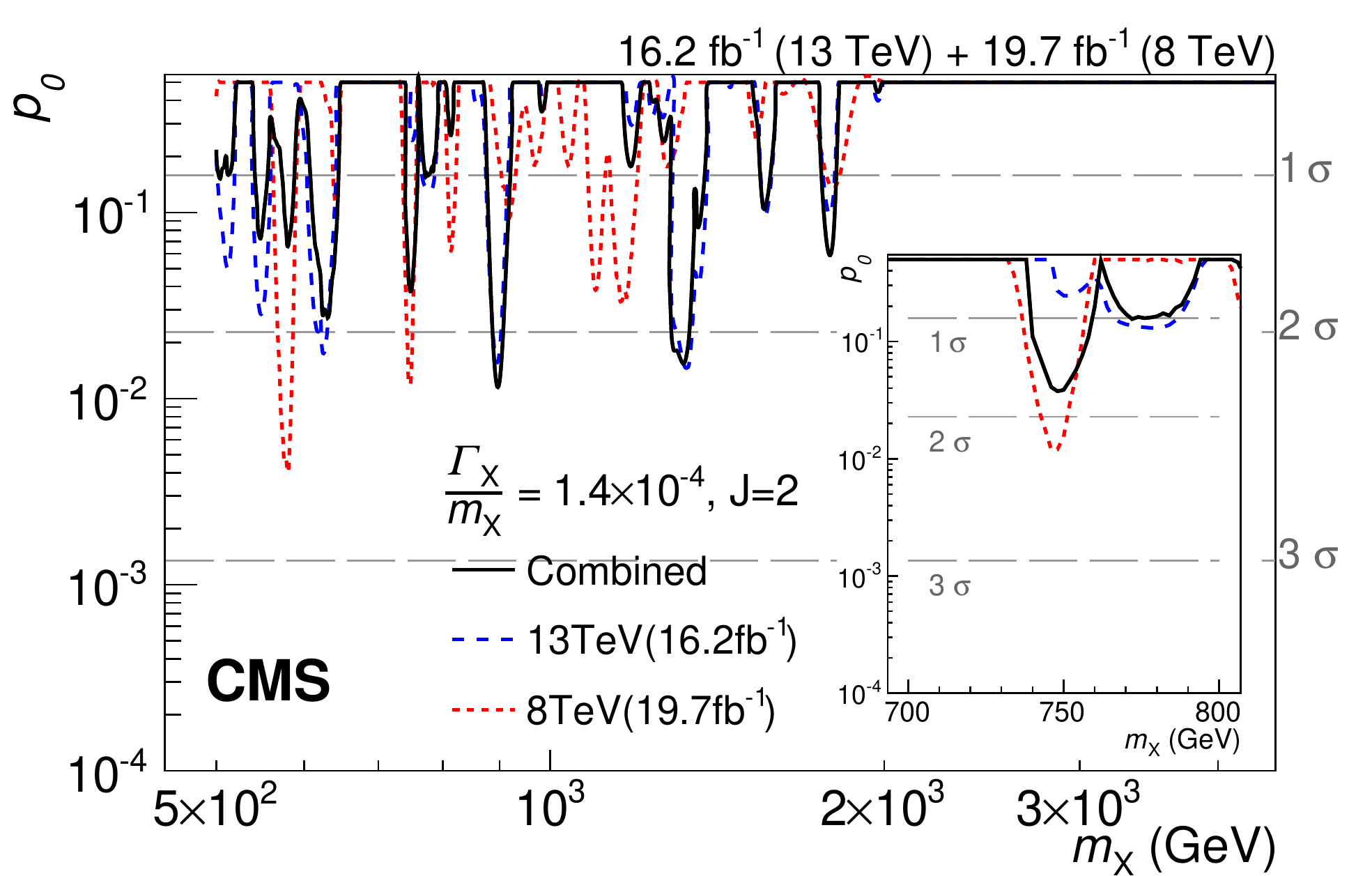}  \\[0.5ex]
    \includegraphics[width=\cmsFigWidth]{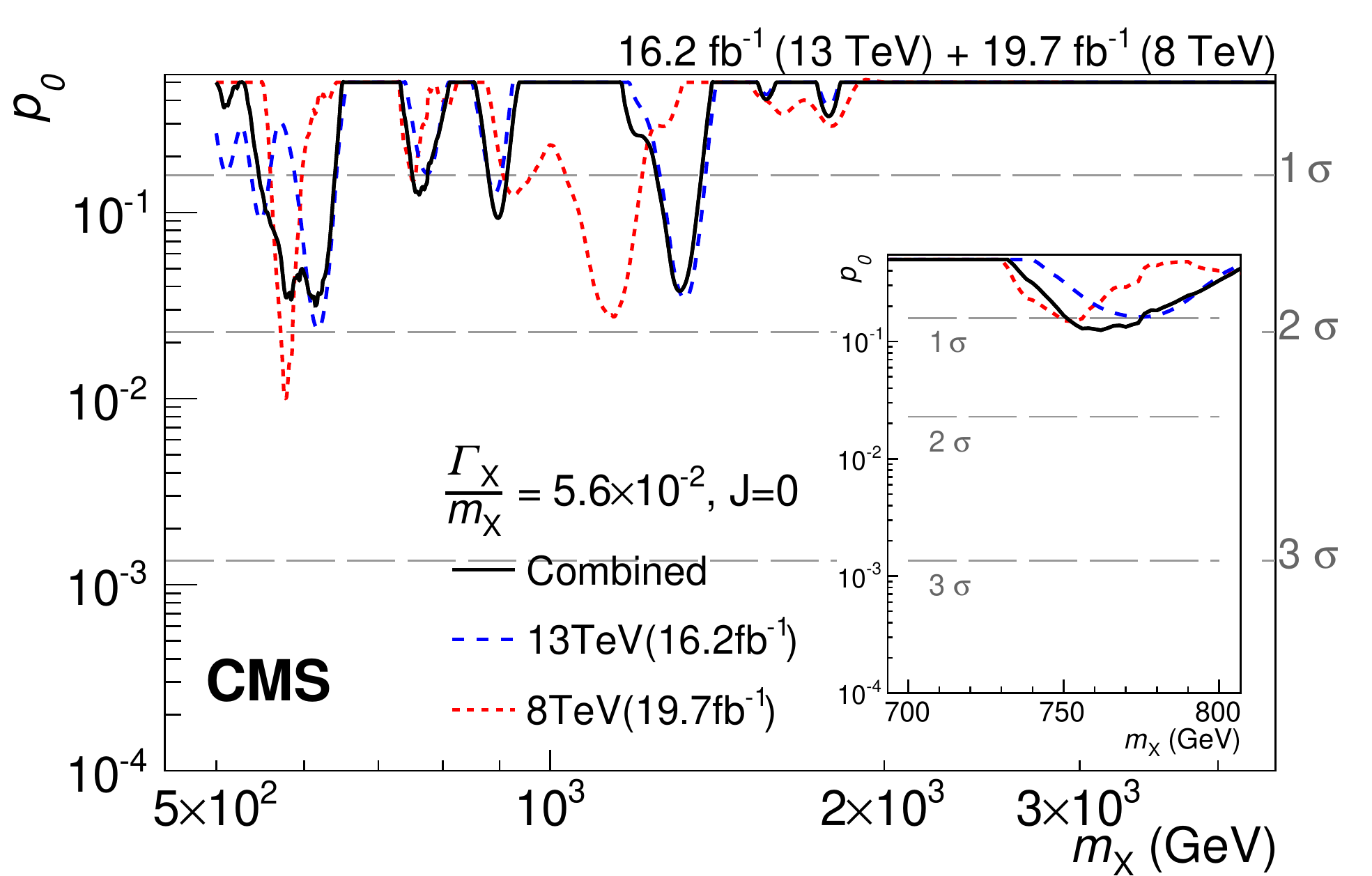}
    \includegraphics[width=\cmsFigWidth]{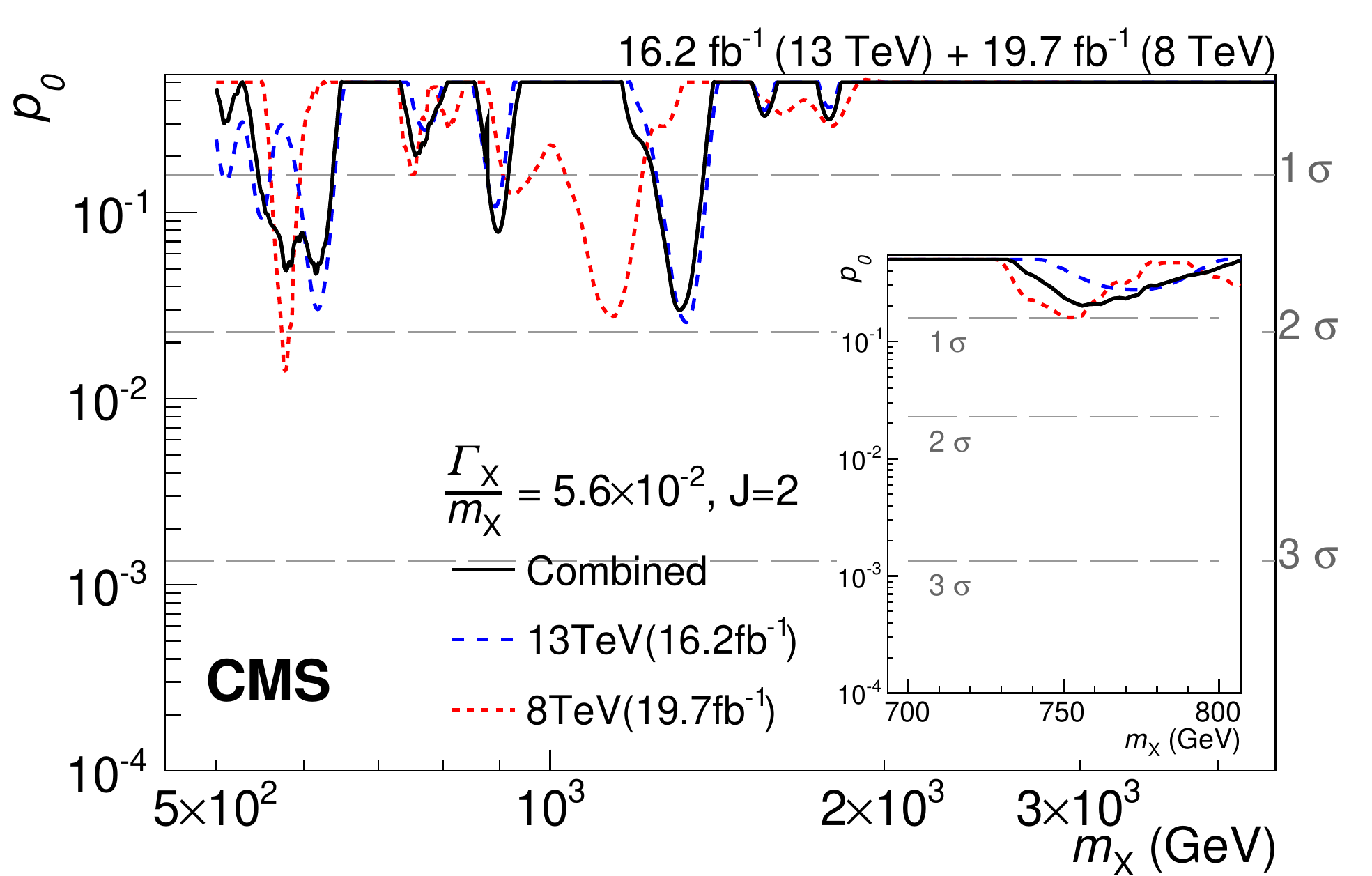}
    \caption{
      Observed background-only $p$-values for resonances with
      (upper) $\GammaOm = 1.4\times 10^{-4}$ and
      (lower) $5.6\times 10^{-2}$ as a function of the resonance mass \mX, from the
      combined analysis of the 8 and 13\TeV data.
      The results obtained for the two individual center-of-mass energies are also shown.
      The curves corresponding to the scalar and \RS graviton hypotheses are shown
      in left and right columns, respectively.
      The insets show an expanded region around $\mX=750\GeV$.
      \label{fig:pvalues:13TeV_8TeV}
    }
\end{figure*}

The observed and expected 95\% CL upper limits on the 13\TeV signal production cross sections
obtained through a combined analysis of the 8\TeV data from 2012 and the 13\TeV data from 2015 and 2016
are shown in Fig.~\ref{fig:limits:13TeV_8TeV}.
Compared to the combined 13\TeV data,
the analysis sensitivity improves by about 10\% at the low end of the \mX range,
while the improvement is negligible at the higher end of the range.
Thus the lower limits on the mass of \RS gravitons
obtained by combining the 8 and 13\TeV data coincide with those obtained with the
13\TeV data alone.

The observed $p_0$ for $\GammaOm = 1.4\times 10^{-4}$ and $5.6\times 10^{-2}$ obtained
with the combined 8 and 13\TeV analysis is shown in
Fig.~\ref{fig:pvalues:13TeV_8TeV}.
The largest excess,
observed for $\mX \approx 0.9\TeV$,
has a local significance of about 2.2 standard deviations,
corresponding to less than 1 standard deviation overall.
For $\mX = 750\GeV$, the excess with 3.4 standard deviation local
significance~\cite{cms-dipho-2015} is reduced to about 1.9 standard deviations.

\section{Summary}

A search for the resonant production of high-mass photon pairs has been presented.
The analysis is based on a sample of proton-proton collisions
collected by the CMS experiment in 2016 at $\sqrts=13\TeV$,
corresponding to an integrated luminosity of~\mylumi.
Events containing two photon candidates with transverse momenta
above 75\GeV are selected.
The diphoton mass spectrum above 500\GeV is examined for evidence of the production
of high-mass spin-0 and spin-2 resonances.

Limits on the production of scalar resonances and
Randall--Sundrum gravitons in the range
${0.5 < \mX < 4.5\TeV}$ and
$1.4\times10^{-4}<\GammaOm < 5.6\times10^{-2}$ are determined
using the modified frequentist approach,
where \mX and \GammaX are the resonance mass and width,
respectively.
The results obtained with the 2016 data set
are combined statistically with those obtained in 2012 and 2015,
corresponding to integrated luminosities of 19.7 and 3.3\fbinv of data
recorded at $\sqrts=8$ and 13\TeV, respectively.

No significant excess is observed above the predictions of the standard model.
Using the leading-order cross sections,
Randall--Sundrum gravitons with masses below 3.85 and 4.45\TeV are excluded
for values of the dimensionless coupling parameter $\ktild=0.1$ and $0.2$, respectively.
For $\ktild=0.01$, graviton masses below 1.95\TeV are excluded,
except for the region between 1.75 and 1.85\TeV.
These are the most stringent limits on Randall--Sundrum graviton production to date.

\begin{acknowledgments}
\bgroup\sloppy
We congratulate our colleagues in the CERN accelerator departments for the excellent performance of the LHC and thank the technical and administrative staffs at CERN and at other CMS institutes for their contributions to the success of the CMS effort. In addition, we gratefully acknowledge the computing centers and personnel of the Worldwide LHC Computing Grid for delivering so effectively the computing infrastructure essential to our analyses. Finally, we acknowledge the enduring support for the construction and operation of the LHC and the CMS detector provided by the following funding agencies: BMWFW and FWF (Austria); FNRS and FWO (Belgium); CNPq, CAPES, FAPERJ, and FAPESP (Brazil); MES (Bulgaria); CERN; CAS, MoST, and NSFC (China); COLCIENCIAS (Colombia); MSES and CSF (Croatia); RPF (Cyprus); SENESCYT (Ecuador); MoER, ERC IUT and ERDF (Estonia); Academy of Finland, MEC, and HIP (Finland); CEA and CNRS/IN2P3 (France); BMBF, DFG, and HGF (Germany); GSRT (Greece); OTKA and NIH (Hungary); DAE and DST (India); IPM (Iran); SFI (Ireland); INFN (Italy); MSIP and NRF (Republic of Korea); LAS (Lithuania); MOE and UM (Malaysia); BUAP, CINVESTAV, CONACYT, LNS, SEP, and UASLP-FAI (Mexico); MBIE (New Zealand); PAEC (Pakistan); MSHE and NSC (Poland); FCT (Portugal); JINR (Dubna); MON, RosAtom, RAS and RFBR (Russia); MESTD (Serbia); SEIDI and CPAN (Spain); Swiss Funding Agencies (Switzerland); MST (Taipei); ThEPCenter, IPST, STAR and NSTDA (Thailand); TUBITAK and TAEK (Turkey); NASU and SFFR (Ukraine); STFC (United Kingdom); DOE and NSF (USA).

Individuals have received support from the Marie-Curie program and the European Research Council and EPLANET (European Union); the Leventis Foundation; the A. P. Sloan Foundation; the Alexander von Humboldt Foundation; the Belgian Federal Science Policy Office; the Fonds pour la Formation \`a la Recherche dans l'Industrie et dans l'Agriculture (FRIA-Belgium); the Agentschap voor Innovatie door Wetenschap en Technologie (IWT-Belgium); the Ministry of Education, Youth and Sports (MEYS) of the Czech Republic; the Council of Science and Industrial Research, India; the HOMING PLUS program of the Foundation for Polish Science, cofinanced from European Union, Regional Development Fund, the Mobility Plus program of the Ministry of Science and Higher Education, the National Science Center (Poland), contracts Harmonia 2014/14/M/ST2/00428, Opus 2013/11/B/ST2/04202, 2014/13/B/ST2/02543 and 2014/15/B/ST2/03998, Sonata-bis 2012/07/E/ST2/01406; the Thalis and Aristeia programs cofinanced by EU-ESF and the Greek NSRF; the National Priorities Research Program by Qatar National Research Fund; the Programa Clar\'in-COFUND del Principado de Asturias; the Rachadapisek Sompot Fund for Postdoctoral Fellowship, Chulalongkorn University and the Chulalongkorn Academic into Its 2nd Century Project Advancement Project (Thailand); and the Welch Foundation, contract C-1845.
\par\egroup
\end{acknowledgments}
\vspace*{3ex}
\bibliography{auto_generated}

\cleardoublepage \appendix\section{The CMS Collaboration \label{app:collab}}\begin{sloppypar}\hyphenpenalty=5000\widowpenalty=500\clubpenalty=5000\textbf{Yerevan Physics Institute,  Yerevan,  Armenia}\\*[0pt]
V.~Khachatryan, A.M.~Sirunyan, A.~Tumasyan
\vskip\cmsinstskip
\textbf{Institut f\"{u}r Hochenergiephysik der OeAW,  Wien,  Austria}\\*[0pt]
W.~Adam, E.~Asilar, T.~Bergauer, J.~Brandstetter, E.~Brondolin, M.~Dragicevic, J.~Er\"{o}, M.~Flechl, M.~Friedl, R.~Fr\"{u}hwirth\cmsAuthorMark{1}, V.M.~Ghete, C.~Hartl, N.~H\"{o}rmann, J.~Hrubec, M.~Jeitler\cmsAuthorMark{1}, A.~K\"{o}nig, I.~Kr\"{a}tschmer, D.~Liko, T.~Matsushita, I.~Mikulec, D.~Rabady, N.~Rad, B.~Rahbaran, H.~Rohringer, J.~Schieck\cmsAuthorMark{1}, J.~Strauss, W.~Waltenberger, C.-E.~Wulz\cmsAuthorMark{1}
% ...
\vskip\cmsinstskip
% ...
1:~~Also at Vienna University of Technology, Vienna, Austria\\
% ...

\end{sloppypar}
\end{document}